\documentclass[aps,prb,twocolumn,groupedaddress]{revtex4}

\usepackage{graphicx}

\begin{document}


\title{Degeneracy, the virial theorem, and stellar collapse}



\author{Christian Y. Cardall}
\email[]{cardallcy@ornl.gov}
\affiliation{Physics Division, Oak Ridge National Laboratory, Oak Ridge,
 	TN 37831-6354}
\affiliation{Department of Physics and Astronomy, University of Tennessee,
	Knoxville, TN 37996-1200} 


\date{\today}

\begin{abstract}
Formulae for the energies of degenerate non-relativistic and ultra-relativistic Fermi gases play multiple roles in simple arguments related to the collapse of a stellar core to a neutron star.  These formulae, deployed in conjunction with the virial theorem and a few other basic physical principles, provide surprisingly good estimates of the temperature, mass, and radius (and therefore also density and entropy) of the core at the onset of collapse; the final radius and composition of the cold compact remnant; and the total energy lost to neutrino emission during collapse.
\end{abstract}


\maketitle

\section{Introduction}

Not long after `supernovae' came to be distinguished from `common novae' by virtue of their greatly higher intrinsic luminosity,\cite{Baade1934On-Super-Novae} and shortly after the discovery of the neutron in the early 1930s, Baade and Zwicky declared:
``With all reserve we advance the view that supernovae represent the
transitions from ordinary stars to {\em neutron stars,} which in their final
stages consist of extremely closely packed neutrons.''\cite{Baade1934Supernovae-and-,Baade1934Cosmic-Rays-fro}  
For the `core-collapse' varieties of supernovae (Types Ib/Ic/II) this basic picture prevails today with good theoretical and observational support.\cite{Arnett1996Supernovae-and-,Woosley2002The-evolution-a,Filippenko1997Optical-Spectra,Lattimer2004The-Physics-of-}  

The virial theorem and formulae for the energies of degenerate Fermi gases are especially important to this paper's simple arguments and calculations related to the collapse of the of a massive star's core to a neutron star. The virial theorem states that a self-gravitating body consisting of particles that are either non-relativistic or ultra-relativistic satisfies
\begin{equation}
2E_{N} + E_{R} = -E_G, \label{virial}
\end{equation}
where $E_{N}$ and $E_{R}$ are the contributions to the body's internal energy in non-relativistic and ultra-relativistic particles respectively, and $E_G$ is the gravitational energy.\cite{Collins-II1978The-Virial-Theo} While hydrostatic equilibrium requires that pressure and density decrease with increasing distance from the center of a spherically symmetric star, an assumption of uniform density allows for simple and surprisingly good estimates. The gravitational energy of a uniform-density sphere of mass $M$ and radius $R$ is
\begin{equation}
E_{G} = -\frac{3}{5}\frac{G M^2}{R}, \label{gravity}
\end{equation}
where $G$ is the gravitational constant. As will be seen below, degenerate gases dominate the internal energies of both the pre-collapse stellar core and the resulting neutron star; hence formulae for the internal energies $(E_F)_{N}$ and $(E_F)_{R}$ of non-relativistic and ultra-relativistic uniform-density degenerate Fermi gases confined to a spherical volume of radius $R$ will play a prominent role in use of Eq. (\ref{virial}). These energies are
\begin{eqnarray}
(E_F)_{N} &=& \frac{3}{5} N_N \,(\epsilon_F)_N, \label{internalEnergyN}\\
(E_F)_{R} &=& \frac{3}{4} N_R \,(\epsilon_F)_R, \label{internalEnergyR}
\end{eqnarray}
where $N_N$ and $N_R$ are the numbers of particles in non-relativistic and ultra-relativistic gases composed of a given particle type, and $(\epsilon_F)_N$ and $(\epsilon_F)_R$ are the respective Fermi energies:
\begin{eqnarray}
(\epsilon_F)_N &=& \left(\frac{9 \pi N_N}{2 g}\right)^{2/3} \frac{\hbar^2}{2m R^2}, \label{fermiEnergyN}\\
(\epsilon_F)_R &=& \left(\frac{9 \pi N_R}{2 g}\right)^{1/3} \frac{\hbar c}{R}. \label{fermiEnergyR}
\end{eqnarray}
Here $g$ is the spin degeneracy of the particle type, $m$ is the particle mass (relevant to the non-relativistic case), $\hbar$ is the reduced Planck constant $h/2\pi$, and $c$ is the speed of light (relevant to the ultra-relativistic case). In addition to their utility in representing the internal energies of self-gravitating bodies, Eqs. (\ref{internalEnergyN})-(\ref{fermiEnergyR}) will also prove useful in a simple model of the nucleus used in arguments about the process of electron capture on nuclei that signals the onset of core collapse.

After the existence of a degenerate iron core at the center of a massive star is argued for in Sec. \ref{sec:ironCore}, subsequent sections utilize Eqs. (\ref{internalEnergyN})-(\ref{fermiEnergyR}) in various ways to determine the properties of this pre-collapse core (Sections \ref{sec:mass} and \ref{sec:radius}) and of the cold neutron star that is the final result of its collapse (Section \ref{sec:remnant}). The energy lost to neutrino emission---which allows collapse to proceed---is addressed in Section \ref{sec:energyLoss}. Section \ref{sec:conclusion} contains concluding remarks.

\section{A degenerate iron core}
\label{sec:ironCore}

Stellar collapse results from the evolution of a massive star.\cite{Arnett1996Supernovae-and-,Woosley2002The-evolution-a} All stars spend most of their lives burning hydrogen to helium in a central core, but in those with mass $M \gtrsim 8\; M_\odot$ (eight times the solar mass) the helium ash eventually ignites and burns to carbon, and then perhaps to oxygen and magnesium. For $M \gtrsim 11\; M_\odot$ additional core burning stages occur with increasing density and temperature, culminating with the production of iron-group nuclei; these lie near the top of the nuclear binding energy curve, and so no further energy can be extracted by nuclear fusion. 

In very rough caricature, burning to higher-mass nuclear species tends to proceed, at least at first, by successive captures of $\alpha$ particles (helium nuclei with proton number $Z=2$ and mass number $A=4$, that is, $_2^4$He). Two factors favor this. First, helium nuclei are unusually tightly bound, and this tends to make their presence in relatively high abundance energetically favorable. Second, their modest proton number makes for a less formidable Coulomb barrier: it is much easier for an $\alpha$ particle to approach a higher-mass nucleus than it is for two higher-mass nuclei to approach each other.

The dominant naturally-occuring isotope of iron is $_{26}^{56}$Fe, whose mass number $A=56$ is indeed an integer multiple of 4. While it is not the most tightly bound nucleus overall,\cite{Fewell1994The-atomic-nucl} it is the most tightly bound nucleus near the valley of $\beta$ stability,\cite{Woosley2002The-evolution-a} which tends towards greater neutron richness with higher $A$ because of the increasingly high Coulomb energy of confined protons. 

The synthesis of $_{26}^{56}$Fe is not merely an unbroken sequence of $\alpha$ captures. A variety of strong and electromagnetic interactions occur; these include fusion of higher-mass species when $\alpha$ particles are scarce, but the high temperatures required by the associated high Coulomb barriers also induce photodissociation that replenishes the $\alpha$ particle abundance, enabling the predominant role of species whose mass numbers are multiples of 4. Moreover, the neutron excess manifest in $_{26}^{56}$Fe requires the conversion of two protons to neutrons for each nucleus of this isotope, and so weak interactions must also occur along the way.

Nevertheless, the assumption that the stellar core burns all the way to $^{56}_{26}$Fe, with a prominent role for $\alpha$ capture, allows the pre-collapse iron core temperature $T_{\rm Fe}$ to be estimated. 

The first step in this estimate is to understand the relationship between temperature and the most favorable energy $E_{\rm Gamow}$ for a nuclear reaction to proceed, which is typically higher than $k T$ (where $k$ is Boltzmann's constant). This most favorable energy is the maximum of the so-called Gamow peak, which represents a compromise between the higher probability of Coulomb barrier penetration for higher energy and the declining number of particles in the high-energy tail of the Maxwell-Boltzmann distribution.\cite{Arnett1996Supernovae-and-} In particular $E_{\rm Gamow}$ is the maximum of the typically narrowly-peaked function $e^{-b/E^{1/2}} e^{-E/kT}$, where $b=\sqrt{2\pi^2 \mu c^2(Z_1 Z_2 \, \alpha)^2}$:
\begin{eqnarray}
E_{\rm Gamow} &=& \left(\frac{b\,kT}{2}\right)^{2/3} \\
&=& \left[\frac{\pi^2\mu c^2 (Z_1 Z_2 \,\alpha\,  k T)^2}{2}\right]^{1/3}.\label{eGamow}
\end{eqnarray}
Here $\alpha \approx 1/137$ is the fine structure constant (the two distinct uses of the symbol $\alpha$---to denote the $_2^4$He nucleus, or the fine structure constant---should be clear from the context); and
\begin{equation}
\mu c^2 = m_N c^2 \frac{A_1 A_2}{A_1+A_2} 
\end{equation}
is the reduced mass of colliding nuclei of proton and mass numbers $Z_1, A_1$ and $Z_2, A_2$, where $m_N c^2 = 931.5$~MeV is the energy of the atomic mass unit.

\begin{figure}
 \includegraphics[width=3in]{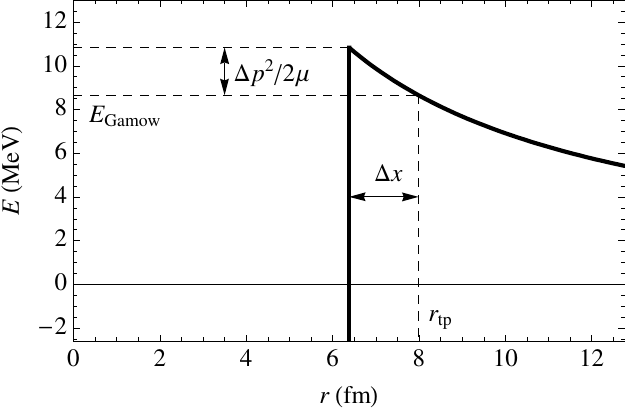}
 \caption{The potential governing a putative final $\alpha$ capture to $_{26}^{56}$Fe (heavy solid line) is taken to be Coulomb repulsion until the colliding nuclei begin to overlap, at which point the strong interaction provides a deep well. The minimum value of $E_{\rm Gamow}$ for which tunneling is effective can be estimated with the uncertainty principle $\Delta x \, \Delta p \approx \hbar$, where $\Delta x$ is the effective width of the barrier (difference between the `turning point' separation $r_{\rm tp}$ and the `overlap' separation), and $\Delta p$ is related to the height of the potential barrier above $E_{\rm Gamow}$ ($\mu$ is the reduced mass of the colliding nuclei).     
 \label{fig:eGamow}}
 \end{figure}

Normally the temperature is considered a `given' and $E_{\rm Gamow}$ is then determined from it; the strategy here will be to estimate the minimum value of $E_{\rm Gamow}$ for which $\alpha$ capture to iron could readily occur, and then invert Eq. (\ref{eGamow}) to determine the corresponding temperature $T_{\rm Fe}$. The uncertainty principle serves to characterize the point at which tunneling through the Coulomb barrier first becomes effective:
\begin{equation}
\Delta x\, \Delta p \, c \approx \hbar  c, \label{uncertainty}
\end{equation}
where the factor of $c$ is included for calculational convenience ($\hbar c = 197.33$ MeV fm). Uncertainties in position $\Delta x$ and momentum $\Delta p$ are associated with the width and height of the barrier as illustrated in Fig. \ref{fig:eGamow}. The width of the barrier is
\begin{equation}
\Delta x = r_{\rm tp} - r_N \left(A_1^{1/3} + A_2^{1/3}\right).
\end{equation}
Here $r_{\rm tp}$ is the classical `turning point' separation at which the Coulomb energy equals $E_{\rm Gamow}$,
\begin{equation}
E_{\rm Gamow} = \frac{Z_1 Z_2 \,\alpha\, \hbar c}{r_{\rm tp}},
\end{equation}
and $r_N A_1^{1/3}$ and $r_N A_2^{1/3}$ are the radii of the colliding nuclei;  when they overlap, the deep energy well provided by the strong interaction dominates the potential governing the collision ($r_N \approx 1.2$~fm is the nucleon radius).\cite{Arnett1996Supernovae-and-} The momentum scale $\Delta p$ associated with the height of the barrier above $E_{\rm Gamow}$ is given by
\begin{equation}
\Delta p \, c = \sqrt{2\, \mu c^2 Z_1 Z_2 \,\alpha \, \hbar c \left[\frac{1}{r_N  \left(A_1^{1/3} + A_2^{1/3}\right)} - \frac{1}{r_{\rm tp}}\right]}. \label{height}
\end{equation}
For a putative final $\alpha$ capture to $_{26}^{56}$Fe one has $Z_1 = 2$, $Z_2 = 24$, $A_1 = 4$, and $A_2 = 52$. With these parameters, Eqs. (\ref{eGamow})-(\ref{height}) yield $r_{\rm tp}=8.0$ fm, $E_{\rm Gamow} = 8.7$ MeV, and $k T_{\rm Fe} = 0.56$ MeV.

By this point the electrons in the iron core must be degenerate, thanks to neutrino emission. The thermal energy 
\begin{equation}
k T_{\rm Fe} \approx 0.56\ \mathrm{MeV} 
\end{equation}
estimated above
is comparable to the electron mass energy $m_e c^2 = 0.511$~MeV, so in the absence of an electron chemical potential of similar magnitude positrons will be thermally produced. But these $e^+$ will annihilate with $e^-$ into $\nu\bar\nu$ pairs that escape freely; hence we expect cooling by $\nu\bar\nu$ emission back down towards $k T\approx m_e c^2 \approx 0.5$~MeV. In response to the resulting loss of pressure support the core contracts to higher density, with the conversion of gravitational potential energy increasing the temperature and bringing about further pair emission, and so on. Continued increase in density with temperature regulated to $\sim 0.5$~MeV/$k$ would eventually result in electron degeneracy, which becomes the dominant source of pressure support. (Iron core central temperatures of $\sim 0.5-1$~MeV/$k$ are in fact seen in stellar evolution codes, though the neutrino pair emission that eventually gives rise to degeneracy actually begins as early as carbon burning,\cite{Arnett1996Supernovae-and-,Woosley2002The-evolution-a} when $e^+ e^-$ pairs begin to be produced from the high-energy tail of the photon distribution.)

\section{Mass of the core}
\label{sec:mass}

Catastrophic collapse ensues when the iron core becomes sufficiently massive that it can no longer be supported against gravity by electron degeneracy pressure. As the mass and density of the iron core increase during the final burning stages, the velocity of the degenerate electrons approaches the speed of light; after this, the electrons' contribution to pressure support cannot be further increased. 

An estimate of the core's maximum mass follows from consideration of a uniform core whose pressure support comes solely from a degenerate Fermi gas of relativistic electrons. The number of electrons in the core is
\begin{equation}
N_e = \frac{M \, Y_e}{m_N}, \label{electronNumber}
\end{equation}
where $M$ now denotes the mass of the core rather than the mass of the entire star, and the electron fraction $Y_e$ is the ratio of the net number density of electrons $n_{e^-}-n_{e^+}$ (or, by charge neutrality, the total number density of protons $n_p$) 
to the total baryon number density $n$. Use of Eq. (\ref{electronNumber}) and $g=2$ in Eqs. (\ref{fermiEnergyR}) and (\ref{internalEnergyR}) for the core's relativistic electron energy $E_R=(E_F)_e$ in Eq. (\ref{virial}), together with $E_N \approx 0$ and $E_G$ given by Eq. (\ref{gravity}), results in
\begin{equation}
\frac{3}{4}  \left(\frac{9 \pi}{4}\right)^{1/3} \left(\frac{M \, Y_e}{m_N}\right)^{4/3} \frac{\hbar c}{R} = \frac{3}{5}\frac{G M^2}{R}. \label{chandrasekhar}
\end{equation}
The cancellation of radius is a clear indication that something goes haywire when a star tries to support itself by relativistic degeneracy pressure alone. 
Equation (\ref{chandrasekhar}) nevertheless yields an estimate of the mass of this extreme configuration, the so-called Chandrasekhar mass:
\begin{equation}
M_\mathrm{Fe} \approx 1.5\: M_\odot \left(Y_{e,\mathrm{Fe}}\over 0.46 \right)^2, \label{mass}
\end{equation}
where $Y_{e,\mathrm{Fe}} \approx 0.46$ is the ratio $Z/A$ of the proton and mass numbers of $^{56}_{26}$Fe. This may be taken as an estimate of the mass of the core upon collapse, and of the compact remnant that results therefrom. (Taking inhomogeneous stellar structure into account yields \cite{Chandrasekhar1984On-stars-their-} $M_\mathrm{Fe} \approx 1.22\: M_\odot \left(Y_{e,\mathrm{Fe}}/ 0.46 \right)^2$.)

\section{Density and entropy per baryon of the core}
\label{sec:radius}

The vigorous onset of electron capture---one proximate cause of the core's instability as it approaches the Chandrasekhar mass---can be used to estimate the density of the iron core just before collapse. This new physical input is needed because, in a uniform-density model, the density follows from knowledge of the radius $R$, which cancelled out of the estimate of the core mass $M$ in the previous section.
Electron capture $e^- + (Z,A) \rightarrow (Z-1,A) + \nu_e$ begins in earnest when the electron Fermi energy
\begin{equation}
(\epsilon_F)_e = \left(\frac{9 \pi}{4} \frac{M \, Y_e}{m_N} \right)^{1/3} \frac{\hbar c}{R} \label{eFermiEnergy}
\end{equation}
rises above the energy threshold required to convert a proton to a neutron---{\em when both of these baryons are bound in a nucleus}.

Hence calculation of the energy threshold for electron capture requires some model of the nucleus; fortunately, the concept of degeneracy once again facilitates a simple estimate.
Crudely considering a nucleus of proton number $Z$ and mass number $A$ as consisting of degenerate Fermi gases of non-relativistic protons and neutrons confined to a volume of radius $r_A = r_N A^{1/3}$, the typical emitted neutrino energy $\bar\epsilon_{\nu_e}$ is not $(\epsilon_F)_e$, nor even $(\epsilon_F)_e - \Delta$, where $\Delta=(m_n-m_p-m_e)c^2 \approx 0.78$~MeV is the difference between the neutron mass energy $m_n c^2$ and the proton and electron mass energies $m_p c^2$ and $m_e c^2$: in addition to $\Delta$, the energy required to lift a bound proton from {\em its} Fermi sea (within the nucleus) to the top of the {\em neutron} Fermi sea (within the nucleus) must first be paid.
Hence $\bar\epsilon_{\nu_e} \approx (\epsilon_F)_e - \Delta - [(\epsilon_F)_{n;A,Z} - (\epsilon_F)_{p;A,Z}]$, where $(\epsilon_F)_{n;A,Z}$ and $(\epsilon_F)_{n;A,Z}$ are the Fermi energies of the neutrons and protons bound in the nucleus. Using Eq. (\ref{fermiEnergyN}) for these Fermi energies results in
\begin{eqnarray}
\bar\epsilon_{\nu_e} \!&\approx&\! (\epsilon_F)_e -\Delta \nonumber \\
& &- \left(9\pi\over 4\right)^{2/3}\!\!\! \frac{\hbar^2}{2 m_N\, r_N^2 } \frac{\left[(A-Z)^{2/3} - Z^{2/3}\right]}{A^{2/3}} \\
\!&\approx&\! (\epsilon_F)_e -\Delta \nonumber \\
& &- \left(9\pi\over 4\right)^{2/3}\!\!\! \frac{\hbar^2}{2 m_N\, r_N^2 } \left[(1-Y_e)^{2/3} - Y_e^{2/3}\right],
\label{neutrinoEnergy}
\end{eqnarray}
where it is assumed that all baryons are locked up in nuclei characterized by the same $Z$ and $A$. 

Electron capture can proceed when the right-hand side of Eq. (\ref{neutrinoEnergy}) is positive; using $(\epsilon_F)_e$ as given in Eq. (\ref{eFermiEnergy}) yields $R_\mathrm{Fe} \approx 8.1\times 10^2$~km at the onset of electron capture for $M_\mathrm{Fe}=1.5\,M_\odot$ and $Y_{e,\mathrm{Fe}} = 0.46$, corresponding to 
\begin{eqnarray}
\rho_\mathrm{Fe} &\approx& 1.3\times 10^{12}\ \mathrm{kg/m}^3
\end{eqnarray}
in the iron core at the onset of collapse.

Knowledge of the density then enables the entropy per baryon of the core to be estimated: 
$s_\mathrm{Fe} \approx 0.91\, k$ receives contributions from iron nuclei of mass number $A=56$, relativistic degenerate electrons, and photons:
\begin{eqnarray} 
s_{A,\mathrm{Fe}} &=& \frac{k}{A}\left\{\ln\left[\frac{m_A}{\rho_\mathrm{Fe}}\left(m_A\, k T_\mathrm{Fe} \over 2\pi \hbar^2 \right)^{3/2}\right] + \frac{5}{2} \right\} \nonumber \\
&\approx& 0.31\, k, \label{sNuclei}\\
s_{e,\mathrm{Fe}} &=& \pi^2\, k\, Y_{e,\mathrm{Fe}} \frac{k T_\mathrm{Fe}}{(\epsilon_F)_{e,\mathrm{Fe}}} \approx 0.58\, k, \\
s_{\gamma,\mathrm{Fe}} &=& \frac{4\pi^2}{45}\frac{k}{(\hbar c)^3}\frac{m_N\,(kT_\mathrm{Fe})^3}{\rho_\mathrm{Fe}} \approx 0.02\, k,
\end{eqnarray}
where $m_A \approx m_N A$ is the mass of a nucleus of mass number $A$, $k T_\mathrm{Fe} \approx 0.56$~MeV as found in Sec. \ref{sec:ironCore}, and $(\epsilon_F)_{e,\mathrm{Fe}} \approx 4.4$~MeV is the electron Fermi energy obtained from Eq. (\ref{eFermiEnergy}). 

(As was the case with $T_\mathrm{Fe}$, the values of $\rho_\mathrm{Fe}$ and $s_\mathrm{Fe}$ obtained here are in reasonable agreement with the central iron core values produced by stellar evolution codes.\cite{Woosley2002The-evolution-a})

\section{Composition and radius of the compact remnant}
\label{sec:remnant}

Having characterized the precollapse core and determined that it is doomed to collapse---and
supposing that the degeneracy pressure of nonrelativistic nucleons will halt collapse---what might we expect the final radius $R_{\rm cold}$ of the compact remnant to be, and why do we expect it to consist of neutrons rather than a mixture of neutrons, protons, and electrons? 

Aware that weak interactions can interchange neutrons and protons, we provisionally take our star of ``extremely closely packed [nucleons]'' to consist of nonrelativistic degenerate Fermi gases of neutrons and protons and a relativistic degenerate Fermi gas of electrons. The energies $(E_F)_n$ and $(E_F)_p$ of the nucleons are
\begin{eqnarray}
(E_F)_n &=& \frac{3}{5} \frac{M\left(1-Y_e\right)}{m_N} \,(\epsilon_F)_n, \label{efn}\\
(E_F)_p &=&  \frac{3}{5} \frac{M \, Y_e}{m_N} \,(\epsilon_F)_p \label{efp}
\end{eqnarray}
according to Eq. (\ref{internalEnergyN}), with 
\begin{eqnarray}
(\epsilon_F)_n &=&\left[\frac{9 \pi}{4}\frac{M\left(1-Y_e\right)}{m_N}\right]^{2/3} \frac{\hbar^2}{2m_N R^2}, \\
(\epsilon_F)_p &=&\left(\frac{9 \pi}{4}\frac{M \, Y_e}{m_N}\right)^{2/3} \frac{\hbar^2}{2m_N R^2} 
\end{eqnarray}
according to Eq. (\ref{fermiEnergyN}).
The virial theorem of Eq. (\ref{virial}), with $E_N = (E_F)_n+(E_F)_p$, 
and $E_R = (E_F)_e$ and $E_G$ as these appear on the left- and right-hand sides of Eq. (\ref{chandrasekhar}) respectively, provides one equation describing the cold compact remnant. 
A second constraint is provided by chemical (`$\beta$' or `weak') equilibrium, which requires 
\begin{equation}
\mu_n = \mu_p + \mu_e,\label{beta}
\end{equation}
where the chemical potentials $\mu_i$ include particle masses. That is, $\mu_i = m_i + (\epsilon_F)_i$ at zero temperature, where $(\epsilon_F)_n$ and $(\epsilon_F)_p$ are given above and 
$(\epsilon_F)_e$ is given by Eq. (\ref{eFermiEnergy}).
Taking $M = M_\mathrm{Fe} = 1.5\: M_\odot$, simultaneous solution of Eqs. (\ref{virial}) and (\ref{beta}) as specified here yields $R_{\rm cold} = 11$ km and $Y_{e,{\rm cold}} = 1.0 \times 10^{-2}$. Because $Y_e$ is the fraction of nucleons that are protons, the label `neutron star' is well deserved indeed! This neutronization occurs (via electron capture) because the increasing density associated with collapse makes the degeneracy energy of relativistic electrons increasingly costly.

Neglecting the presence of protons and electrons---that is, taking $Y_e \approx 0$---the virial theorem becomes $(E_F)_n  = -E_{G}/2$, and this alone yields 
\begin{equation}
R_{\rm cold} \approx 11 {\rm\; km} \left(1.5\; M_\odot \over M \right)^{1/3}. 
\end{equation} 
(The resulting baryon number density $n \approx 0.25\; {\rm fm}^{-3}$ is within a factor of two of the value $(4\pi r_N^3 / 3)^{-1} \approx 0.14\; {\rm fm}^{-3}$ derived from from experimental measurements of the size of atomic nuclei, which also feature densely-packed nucleons. In some respects, a neutron star is not unlike an atomic nucleus of astronomical proportions!)

\section{Energy loss to neutrinos}
\label{sec:energyLoss}

The vigorous onset of electron capture not only signals the onset of collapse, as discussed in Sec. \ref{sec:radius}; the energy loss by neutrino emission incident to continued electron capture is necessary for collapse to continue to completion.
This is another consequence of the virial theorem: from Eq. (\ref{virial}) we see that the total energy of the precollapse core, whose internal energy is completely dominated by relativistic electrons, is $(E_F)_e + E_{G} \approx 0$; while the total energy of the cold neutron star, dominated by nonrelativistic neutrons, is $(E_F)_n + E_{G} \approx E_{G}/2$.
Hence an amount of energy equal in magnitude to half the final neutron star gravitational energy,
\begin{equation}
E_{\nu} \approx 1.6\times 10^{46}\;\mathrm{J}\left(M \over 1.5\; M_\odot\right)^{2}\left(11{\rm\; km}\over R\right), \label{energyRelease}
\end{equation} 
must be lost to neutrinos (and a bit of it to the supernova explosion) before the neutron star can reach its cold final state.

\section{Conclusion}
\label{sec:conclusion}

The observed supernova phenomenon, which led Baade and Zwicky to speculate on the formation of neutron stars, exhibits explosion energies of about $10^{44}$~J---a tiny fraction of the $10^{46}$~J of gravitational potential energy released in stellar collapse. Because the explosion is a comparatively minor detail in energetic terms, it is not surprising that sophisticated simulations are required to understand it.\cite{Mezzacappa2005ASCERTAINING-TH,Woosley2005The-physics-of-} 

Nevertheless, some features of the stellar core collapse itself can be understood very simply, using estimates that make use of the virial theorem and formulae for the energies of degenerate gases. (More detailed but less self-contained studies of the nuclear physics, weak interactions, and hydrodynamics related to the state of the pre-collapse core and the process of collapse, which retain an analytic or semi-analytic flavor, are available in the literature.\cite{Bethe1979Equation-of-sta,Fuller1982Stellar-weak-in,Fuller1985Stellar-weak-in,Yahil1983Self-similar-st}) The virial theorem, which relates internal energy to gravitational energy, is a concise summary of the hydrostatic equilibrium that determines stellar structure; an assumption of uniform density then provides simple approximate relations between the mass and radius of both the pre-collapse stellar core and the post-collapse neutron star. The pre-collapse core is supported by electron degeneracy pressure, and when these electrons are relativistic an estimate of the core mass just before collapse is obtained. The radius (and therefore density) of the pre-collapse core can be obtained by estimating conditions for the onset of electron capture, through which energy (and pressure support) are lost through neutrino emission. This calculation involves another use of the concept of degeneracy, namely to crudely model the nucleus as consisting of degenerate proton and neutron ideal gases. Taking the cold post-collapse neutron star to be supported by degenerate non-relativistic nucleons, the virial theorem and the condition of `chemical' equilibrium provide an estimate of the neutron star radius and demonstrate why the name `neutron star' is well-deserved. Finally, the difference in total energy of the pre-collapse core and the post-collapse neutron star implied by virial theorem constitutes an estimate of the energy loss to neutrino emission that allows collapse to occur.


\begin{acknowledgments}
This work was supported by Oak Ridge National Laboratory (ORNL), managed by UT-Battelle, LLC, for the DoE under contract DE-AC05-00OR22725. 
\end{acknowledgments}

\def\rmp{Rev. Mod. Phys.}
\def\apj{Astrophys. J.}


\end{document}